\documentclass[11pt,oneside,a4paper]{article}

\usepackage{graphicx,color,amsmath}
\usepackage{hyperref}
\usepackage[normalem]{ulem}
\usepackage{natbib}
\usepackage{apalike}

		\title{A higher-derivative model predicts the smoothness–duration relationship in head-pointing movements}

	\author{N. Boulanger$^1$, F. Buisseret$^{2,3}$, J. Burny$^{2}$, F. Dierick$^{2,4,5}$, \\ W. Estievenart$^{2}$, A. Teregulov$^{2}$, O. White$^6$ }
	
	\date{%
		\textit{$^1$ Service de Physique de l'Univers, Champs et Gravitation, Universit\'{e} de Mons, UMONS  Research Institute for Complex Systems, Place du Parc 20, 7000 Mons, Belgium. \\ $^2$ Laboratoire Forme et Fonctionnement Humain (FfH), CeREF -- Sant\'e, Haute Ecole Louvain en Hainaut, 136 Rue Trieu Kaisin, Montignies-sur-Sambre, 6061, Belgium. \\ $^3$ Service de Physique Nucl\'{e}aire et Subnucl\'{e}aire, Universit\'{e} de Mons, UMONS Research Institute for Complex Systems, 20 Place du Parc, 7000 Mons, Belgium. \\ $^4$ Facult\'{e} des Sciences de la Motricit\'e, Universit\'e catholique de Louvain, 1 Place Pierre de Coubertin, 1348 Louvain-la-Neuve, Belgium. \\$^5$ Centre National de R\'{e}\'{e}ducation Fonctionnelle et de R\'{e}adaptation -- Rehazenter, RehaLAB, Luxembourg, Grand-Duch\'{e} de Luxembourg.\\$^6$ INSERM UMR1093-CAPS, Université Bourgogne Europe, UFR des Sciences du Sport, F-21000 Dijon, France. } \\
	e-mail: nicolas.boulanger@umons.ac.be, buisseretf@helha.be, la219983@student.helha.be, estievenartw@helha.be, frederic.dierick@gmail.com, la218222@student.helha.be, olivier.white@u-bourgogne.fr}
	
\begin{document}
\maketitle
\begin{abstract}
Head movements require integration of visual, proprioceptive, vestibular, and cervical motor signals. The dynamical principles governing their temporal organization remain unclear. We tested whether a Pais-Uhlenbeck oscillator -- a higher-derivative model with one internal frequency parameter $\omega$ -- can model head kinematics during virtual-reality-based head pointing. Our model predicts a parabolic relationship between movement duration ($T$) and smoothness (log-dimensionless jerk, $LDLJ$) that we have experimentally checked. Sixty-two healthy young adults performed horizontal and vertical head movements with an amplitude of 30$^{\rm o}$. Movement duration, endpoint accuracy, peak angular velocity, overshoot, and $LDLJ$ were extracted from headset kinematics. A mixed-effects parabolic regression confirmed the predicted $LDLJ$ vs $T$ parabolic relation ($R^2 = 0.851$). The quadratic coefficient was not significantly modified by direction or participant, but the intercept and linear coefficient differed between horizontal and vertical movements. As an internal consistency check of the model, we find that the parameter $\omega$, as estimated from the regressions, gives time durations that closely match the measured ones. These findings outline a nonlinear temporal organization of head pointing modulated by direction-specific biomechanical constraints, and suggest that higher-derivative mechanics may provide a principled, non-invasive framework for quantifying cervical motor planning, warranting further validation in clinical populations.

\end{abstract}

\section{Introduction}

Goal-directed movements are conventionally described through movement time and binary success. However, reaching a target rapidly is not equivalent to reaching it well. In a pointing task, success primarily means achieving acceptable spatial accuracy at the endpoint targets: For example, a rotational movement is successful if the angular error between the final pointer position and the target centre falls within an accepted tolerance zone. Other parameters such as movement time and smoothness are better understood as descriptors of the control strategy used to achieve this accuracy. This framing motivates the present study.

The head-neck system is a particularly relevant model for studying motor performance because head movements require the integration of visual, vestibular, proprioceptive, and cervical motor signals. Cervical mobility has been assessed with virtual-reality (VR) methods in both healthy individuals and patients with neck pain, with previous work demonstrating reliable quantification of neck kinematics and movement smoothness \citep{bahat10,bahat16}. One of the advantages of VR is its measurement accuracy through Inertial Measurement Unites (IMUs). For example, the Oculus Quest 2 VR headset is equipped with numerous sensors that enable precise assessment of kinematic parameters, which were analyzed in our study. VR has previously demonstrated its usefulness in assessing cervical mobility. \cite{bahat16} highlight its high reliability in the analysis of cervical kinematics, particularly for average and maximum velocities, which exhibit very good measurement reproducibility. 

The DidRen VR task, presented by \cite{DIERICK2024103270}, is well suited to extend this approach: it requires rapid and accurate head rotations toward visual targets, and the VR implementation allows target size, distance, validation time, and movement direction to be manipulated in a controlled manner while head kinematics are recorded through the VR headset. Previous work confirmed that the DidRen VR task satisfies Fitts's law \citep{fitts1954information}, establishing it as a genuine speed-accuracy paradigm \citep{DIERICK2024103270}.

Fitts' law, however, remains an empirical speed-accuracy relation. It does not explain how the temporal structure of movement is organized, nor how this structure depends on smoothness, gravitational load, or movement direction. A broader theoretical context is provided by stochastic and optimal feedback control (OFC), in which the nervous system selects motor commands that satisfy task-relevant endpoint constraints while balancing costs related to effort, variability, duration, and smoothness \citep{harris98,todorov02}. Fitts' law can be seen as one behavioural expression of such a control problem, rather than as the underlying theory. 

Jerk has a central but often misunderstood role in this debate. In the classical minimum-jerk model, smooth arm trajectories emerge from minimizing the time integral of squared jerk \citep{Flash1688}. Our approach is distinct: we do not assume that the motor system directly minimizes squared jerk. Instead, we propose a higher-derivative (HD) model in which jerk is a dynamical state variable that participates in endpoint realization. Initial jerk, and not initial acceleration, is the critical planning variable in this framework, consistent with recent HD formulations of motor control and neural dynamics \citep{physics6040077,frontiers}. Formally, pointing is treated as a higher-derivative dynamical problem subject to endpoint position and velocity constraints. As we show in the present note, the simplest admissible action, of Pais-Uhlenbeck form \citep{Pais:1950za}, yields a testable parabolic prediction linking log-dimensionless jerk ($LDLJ$), a validated dimensionless smoothness index \citep{bala_smooth}, to movement duration $T$. The main goal of the present study was to test this prediction in horizontal and vertical head pointing movements recorded during the DidRen VR test. We hypothesized that both directions would obey this common parabolic law, reflected in a shared quadratic coefficient. We further hypothesized that horizontal and vertical movements would differ in their intercept and linear slope, reflecting direction-dependent differences in the cost of producing smooth head movements. A secondary goal of this study was to compare movement indices such as duration, accuracy, and maximal rotation speed, between the two rotations analyzed, i.e. vertical (pitch) and horizontal (yaw).

\section{Material and methods }
\subsection{A higher-derivative model for pointing tasks}

From our works \cite{physics6040077,frontiers} one can deduce that the simplest action that is compatible with minimal requirements about motor control is a $N=2$ higher-derivative one:
\begin{equation}
	{\mathcal S}[x]=\int^T_0 L(x,\dot x,\ddot x)\, dt.
\end{equation}
For a pointing task, i.e. a voluntary movement going from an initial to a final position with zero initial and final speed, one may impose 4 initial conditions : 
\begin{equation}\label{initial}
	x(0)=0,\ \dot x(0)=0,\ \ddot x(0)=a,\ \dddot x(0)=j,
\end{equation}
with $a$ and $j$ the initial acceleration and jerk, respectively. We assume in our analysis that $j>0$. The target is reached after a time $T$ when the velocity reaches zero, with a position $x(T)$ equal to the target's one -- up to some tolerance. The relation
\begin{equation}\label{final}
	\dot x(T)=0
\end{equation}
has thus also to be imposed. 

Still according to \cite{physics6040077}, the simplest action, that we use as a toy model, is of Pais-Uhlenbeck type \citep{Pais:1950za}:
\begin{equation}\label{SPU}
	{\mathcal S}_{PU}[x]=\int^T_0\left(\frac{\ddot x^2}{2}-\frac{\omega^2}{2}\dot x^2\right)dt.
\end{equation}
It depends on a single parameter $\omega$, that we call ``internal frequency". 
Setting to zero the variation of the action ${\mathcal S}_{PU}$ 
under the constraints that the position $x(t)$ and velocity $\dot{x}(t)$ 
should be fixed at $t=0$ and $t=T$ leads to the 
equations of motion 
\begin{equation}\label{eom}
	x^{(4)}+\omega^2\ddot x=0\,.
\end{equation}
Multiplying it by $\dddot x$ shows that there exists a conserved quantity
\begin{equation}\label{Invariant}
	{\mathcal E}=\frac{\dddot x^2}{2}+\frac{\omega^2}{2}\ddot x^2
\end{equation}
linking jerk and acceleration. 
The general solution to (\ref{eom}) is given by
\begin{equation}\label{solu0}
	x(t)=K_1\sin\omega t+K_2\cos\omega t+K_3 t+K_4,
\end{equation}
with the $K_i$ being four real constants. Imposing the initial conditions (\ref{initial}) to (\ref{solu0}), one has
\begin{equation}
	x(t)=-\frac{j}{\omega^3}\sin\omega t-\frac{a}{\omega^2}\cos\omega t+\frac{j}{\omega^2} t+\frac{a}{\omega^2}.
\end{equation}

Now we define 
\begin{equation}\label{phidef}
	\tan\phi=\frac{a\omega}{j}, \qquad x_A=\frac{2\pi\, j}{\omega^3},
\end{equation}
where $\phi\in\left[-\frac{\pi}{2},\frac{\pi}{2}\right]$, and rewrite the above solution as
\begin{equation}\label{solu1}
	x(t)=\frac{x_A}{2\pi\cos\phi}\left(-\sin(\omega\, t+\phi)+\omega\, t\, \cos\phi+\sin\phi\right).
\end{equation}
Note that the invariant (\ref{Invariant}) takes the value 
\begin{equation}
\label{varEpsilon}
{\mathcal E}=\frac{j^2}{2\cos^2\!\phi}=\frac{j^2+a^2\omega^2}{2}\;,
\end{equation}
on the solution.

The equation $\dot x(t_0)=0$, that leads to the moments at which speed vanishes, has a simple form in terms of $\phi$: $\cos(\omega t+\phi)=\cos\phi$. 
The first roots are: $\omega\, t_0=0$, $2\pi$, 
together with 
$\omega\, t_0 = -2\phi$ if $\phi<0$, and $\omega\, t_0 = 2\pi-2\phi$ if $\phi>0$. 
The movement duration is given by the largest value of $t_0$, that is
\begin{equation}\label{Tdef}
	T_0=\frac{2\pi}{\omega}, 
\end{equation}
such that the final position is estimated by
\begin{equation}
	x(T_0)=x_A,
\end{equation}
coinciding with the target position if $j$ is well chosen by the individual performing the movement. If $\phi>0$, there is overshooting, i.e. a maximal position reached at $t_M=\frac{2\pi}{\omega}-\frac{2\phi}{\omega}$ such that 
\begin{equation}\label{osdef}
	\delta_+=x_M-x_A=x(t_M)-x_A=\frac{\tan\phi - \phi}{\pi}>0.
\end{equation}
The quantity $\delta_+$ assesses the overshoot magnitude. 

When $\phi < 0$ (which corresponds to an initial negative acceleration $a < 0$), the model predicts a trajectory that initially moves away from the target, reaching a local minimum at $x(-\frac{2\phi}{\omega}) < 0$. Because this represents an initial movement in the opposite direction to the goal, we consider this specific kinematic case physiologically irrelevant for a direct, goal-directed pointing task.

\subsection{Population}

Participants were students from Physical Therapy or Occupational Therapy Departments at Haute Ecole Louvain en Hainaut. A written informed consent was signed by each participant after being informed of the experimental protocol of the study and the use of their personal data. The experimental protocol has been accepted by the local internal commission and respects the Helsinki Declaration on Ethical Principles for Medical Research Involving Human Beings. The study protocol was approved by the Academic Ethical Committee of Brussels Alliance for Research and Higher Education (Protocol B-200-2025-071). 

Participants had to: be aged between 18 and 25 years; have a Neck Disability Index (NDI) $\leq$ 8$\%$ (no disability according to \cite{VERNON2008491}); have not consumed alcohol or drugs within 24 hours prior to the test; have not experienced neck injury or neck surgery within the past 6 months. Participants were excluded if they felt cybersickness in our VR environment (we refer the interested reader to \cite{martirosov} for a detailed discussion of this concept).

\subsection{Protocol}

Participants sat on a chair with a backrest but no armrests and equipped by the experimenters with the VR headset Oculus Quest 2 (Oculus VR Inc., Irvine, CA, USA), connected to a computer. Participants were placed in the DidRen VR setup; we refer the reader to \cite{DIERICK2024103270} for a detailed description of the latter VR application and report here the key points. 

The DidRen VR setup is an environment in which participant's head movements control a blue circle called the ``laser beam". The pointing task consists in reaching successive targets with the laser beam. The instruction given to each participant was ``When a target becomes red, you must reach it with the beam as accurately as possible, keep it in the target until it is validated with audible confirmation, then move on to the next one". Two target setups were used, defining the two tests  realized by the participants: horizontal and vertical in randomized order. In the horizontal setup, 3 horizontal targets (center-left-right) are separated by 30$^{\rm o}$. One complete horizontal test consists of five cycles of a 4-movement sequence per target distance (angle of rotation); the sequence being predefined (center-right-center-left-center). In the vertical setup, 3 vertical targets (center-up-down) are separated by 30$^{\rm o}$. One completed vertical test consists of five cycles of a 4-movements sequence per target distance (angle of rotation); the sequence being predefined (center-up-center-down-center). The validation time, i.e. the time during which the center of the laser beam must stay inside the validation zone of the target for the target to be validated, was set equal to 0.8 s. The validation zone is 2$^{\rm o}$ wide. In the considered task, the accuracy $\delta$ is defined as the angular endpoint error between the laser-beam position and the geometric center of the validated target at the moment of auditory confirmation. A movement is successful if $\delta$ falls within the accepted tolerance zone, i.e. 2$^{\rm o}$.

In order to become familiar with the task, a practice horizontal test was first carried out before measurements during which instructions were given by the experimenter. Head angular velocity, $\vec v$ was recorded versus time at a sampling frequency of 100 Hz for each target direction. The total time of the test with targets in horizontal position was also recorded; it is the DidRen time measured in the study of \cite{hage19} among other parameters. The VR headset frame is denoted X (pitch), Y (yaw), Z (roll). Positive (negative) angular speed along the X axis denote extension (flexion) movement. Positive (negative) angular speed along the Y axis denote left (right) rotation movement. Positive (negative) angular speed along the Z axis denote right (left) inclination movement.

\subsection{Data analysis }

\begin{figure}[h]
	\begin{center}
		\includegraphics[width=0.8\linewidth]{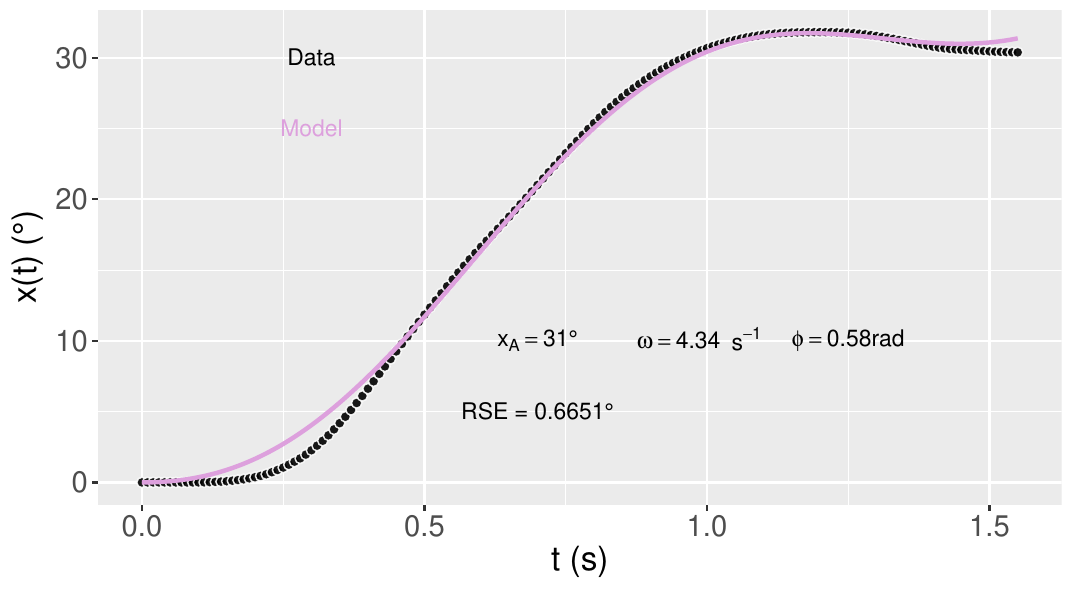}
	\end{center}
	\caption{Typical trace of a movement recorded during the DidRen VR test. The black points are the recorded rotation angles during a pointing task from the central target to the right target. The solid line shows model (\ref{solu1}) fitted on the points by a least square method. The fit has been performed using the \textit{nls} function in R software and the optimal parameters are given in the plot, together with the Residual Standard Error (RSE).}
	\label{fig:typical}
\end{figure}

First, we computed the rotation angles along the three rotation axis for each test, $\vec x$, by integrating $\vec v$ with zero initial values -- $\vec v$ was not filtered. The integration was numerically performed by using the trapezoidal rule. The relevant angular degree of freedom was kept, X or Y for vertical or horizontal target setup respectively, and denoted $x$ in the following. Each movement was then identified, the start being when the laser beam gets out of the first target's validation zone and the ending being when the beam enters the second target's validation zone. A typical trace of a movement with overshoot can be appraised in Fig. \ref{fig:typical}, 
where the close agreement with our model (\ref{solu1}) is also displayed. 
The transformation $x(t)\mapsto \vert x(t)-x(0)\vert$ was then applied to each movement to remove sign differences due to the movement direction and simplify the numerical analyses.

Second, we computed several indices to assess each movement: $T$, accuracy $\delta$, overshoot $\delta_+$, and absolute maximal speed $v_M$. The last indicator assessing the motor strategy is kinematical smoothness. A robust indicator of smoothness is the log-dimensionless jerk \cite{bala_smooth}, reading, in our notations,
\begin{equation}\label{LDLJdef}
	LDLJ=-\ln\left(\frac{T^3}{v_M^2}2\pi\, I(T)\right),
\end{equation}
where $I$ is the mean-squared jerk integral defined below, see the equation 
\eqref{It} of the appendix. The jerk is computed by differentiating twice 
the angular velocity, and the integral is again computed by resorting 
to trapezoidal rule. \cite{SIMONSEN2025112448} have shown that a sampling frequency of 84 Hz was enough to reach reliable values of jerk integrals. This suggests that the 100 Hz sampling frequency is compatible with reliable jerk-integral estimation, although the influence of filtering choices should be assessed in future work.

We may compare $LDLJ$ to its theoretical value within our pointing task 
model. As shown in Appendix, $LDLJ$ as a function of $T$ is expected to have a parabolic 
shape, 
\begin{equation}\label{model_eq}
	LDLJ= C_0 -C_1\ T+C_2\ T^2,	
\end{equation}
where the coefficients are given in Appendix for small values of $\phi$. We stress that the above $LDLJ$ vs $T$ relationship should not be interpreted as a purely empirical correlation between independent variables. Rather, the test concerns whether the empirically observed dependence follows the specific parabolic form predicted by our higher-derivative model.

Because equation (\ref{model_eq}) is used as a local approximation around $T_0$, the fitted intercept $C_0$ should not be interpreted as a theoretical value of $LDLJ$ at $T=0$. The relevant behaviours to be looked for are rather given by $C_1$ and $C_2$: In the small-$\phi$ regime, a leading-order estimate of the internal frequency can be obtained by setting $\phi=0$, leading to
\begin{equation}\label{omegadef}
	\omega=\frac{48\pi}{7}\frac{C_2}{C_1}\;,
\end{equation}
see Eqs. (\ref{C1def}) and (\ref{C2def}).

The kinematical data recorded by the VR headset were analysed as described in this section using a homemade Python code with the following libraries: \textit{numpy}, \textit{pandas}, \textit{pathlib}.

\subsection{Statistical analysis}

We first compared the values of $T$, $\delta$, $\delta_+$, $v_M$ and $LDLJ$ in the vertical and horizontal directions, averaged on all the movements by participant. The comparison was made by using a paired t-test if normality was not rejected by a Shapiro-Wilk test, or by using a Wilcoxon test if normality was rejected. Paired effect sizes were computed: Cohen's $d$  in case of a paired t-test, and Cohen's $r$ in case of a Wilcoxon test \citep{cohenNP}.

Then we fitted a linear mixed-effects model with $LDLJ$ as  dependent variable, fixed effects of $T$, $T^2$, direction, $T\times$ direction, and $T^2\times$ direction. Participant-specific random intercepts,  $T\times$participant and $T^2\times$participant effects were added. Participant$\times$Direction effects were neglected. The fit was performed on all the movements performed by all participants. Calculations were made with a significance threshold of 0.05 using JMP Student Edition 19. 

\section{Results}
\subsection{Population}

Sixty-seven participants were included in the study and five were excluded; the characteristics of the 62 participants having completed the protocol are presented in Table \ref{tab:pop}. Data recording was incomplete for 3 participants and 2 participants refused to participate in all the conditions (lack of time). For each participant and condition, 5 cycles of 4 movements were recorded, yielding 20 movements per participant per condition. With 62 participants, this corresponded to 1240 movements per condition. 

\subsection{Impact of the direction}

The differences induced by the direction of the targets can be appraised in Table \ref{tab:direc}. In the horizontal direction, movement shows a significantly higher $v_M$ and a higher smoothness ($LDLJ$) 
but also a higher overshoot  ($\delta_+$) than in the vertical direction. 

\subsection{Smoothness-duration relationship}

\begin{figure}[h]
	\centering
	\includegraphics[width=0.8\linewidth]{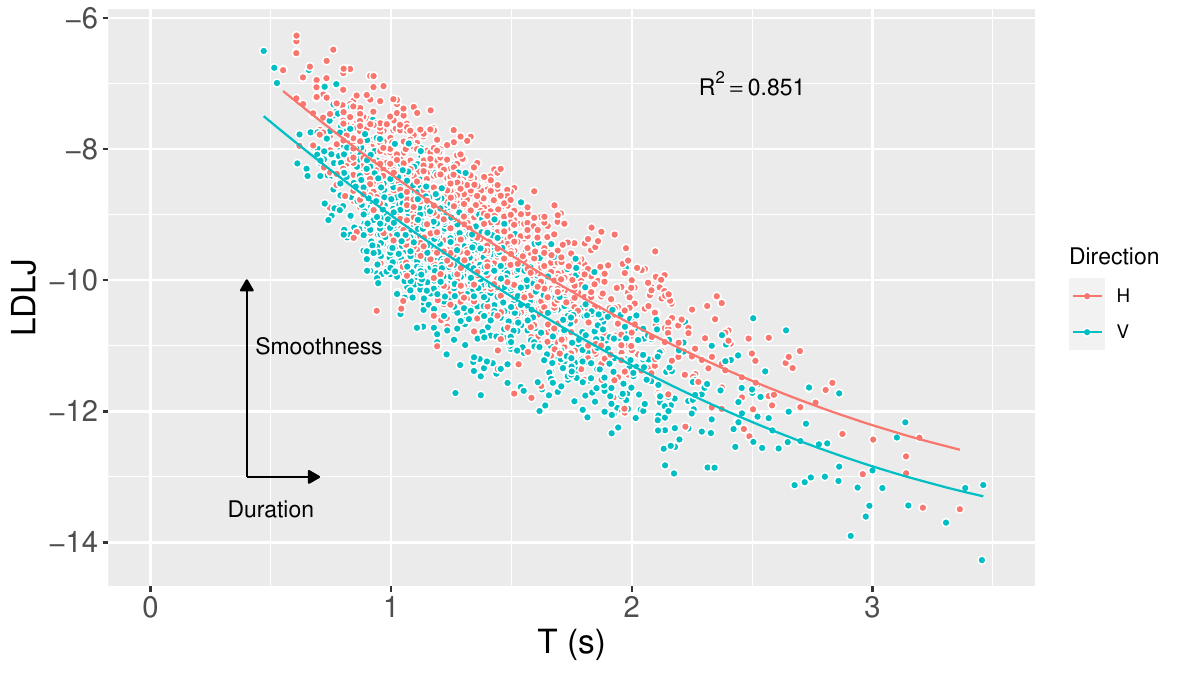}
	\caption{Plot of $LDLJ$ versus $T$ during movements from one target to another during DidRen VR tests in our population. Red and light blue points are movements in the horizontal (H, yaw movements) and vertical  (V, pitch movements) directions respectively. The parabolas (lines) are the best fits according to our mixed model regressions, averaged on all participants; the model's coefficient of determination is indicated to appraise the quality of the models.}
	\label{fig:fit}
\end{figure}

The parabolic mixed model regression explained a large proportion of variance with a $R^2$ of 0.851, as it can be graphically appraised in Fig. \ref{fig:fit}. The intercept is nonzero ($p<0.0001$); the factors participant and direction significantly modify the intercept ($p<0.0001$ for both factors): $C_0=-6.105\pm 0.070$ in the vertical direction and $-5.788\pm0.071$ in the horizontal direction. The linear term is nonzero ($p<0.001$); the factor direction significantly modifies this term but ($p=0.002$) but not the factor participant ($p=0.105$): $C_1=2.601\pm 0.036$ s$^{-1}$ in the vertical direction and $2.522\pm 0.044$ s$^{-1}$ in the horizontal direction. The quadratic term is nonzero ($p<0.001$); the factors direction and participant do not significantly modify this term ($p=0.545$ and $p=0.266$): $C_2=0.557\pm 0.040$ s$^{-2}$ in the vertical direction and $0.538\pm 0.051$ s$^{-2}$ in the horizontal direction. 

Using the leading-order estimate in equation (\ref{omegadef}), we obtained $\omega=4.365$ s$^{-1}$ in the vertical direction and 4.595 s$^{-1}$ in the horizontal direction. These estimates allow a parameter-free consistency check: the predicted ideal movement durations $\frac{2\pi}{\omega}=\frac{14C_1}{48 C_2}$ equals 1.439 s (vertical) and 1.367 s (horizontal), which are compatible with the observed distributions of $T$ reported in Table \ref{tab:direc}. This suggests that an internal frequency-like parameter can be estimated from behavioural kinematics and may provide a compact descriptor of temporal movement organization.

\section{Discussion}

The present study shows that head pointing movements in the DidRen VR test exhibit a systematic parabolic relationship between movement duration and smoothness. This extends the interpretive scope beyond Fitts' law. Fitts' law validates the task as a speed-accuracy paradigm: target difficulty modulates movement time. Even if a kind of speed-accuracy tradeoff may be observed in our results in the difference between horizontal (larger maximal speed but more overshoot) and vertical (lower maximal speed but less overshoot) rotations, our analysis addresses a different question: how is movement temporally organized once the participant attempts to reach the target? Smoothness, assessed by $LDLJ$, is an indicator of movement's temporal organization. It is higher on average during horizontal rotations than vertical rotations. Moreover, our HD model answers this by linking $LDLJ$ to $T$ through a parabolic law whose coefficients carry distinct motor-control meanings.

The mixed-model parabolic regression explained a large proportion of variance in $LDLJ$ ($R^2 = 0.851$). An important consistency check follows from equation (\ref{omegadef}) of the model, which allows the internal frequency parameter $\omega$ to be estimated directly from the regression coefficients: $\omega = 4.365$ s$^{-1}$ for vertical and 4.595 s$^{-1}$ for horizontal movements. These values predict ideal movement durations of $2\pi/\omega$  which are compatible with the distributions observed in Table \ref{tab:direc}. This provides an internal consistency check of the model: a physiological control parameter is recovered from behavioural kinematics alone, without neuromuscular instrumentation.

The three coefficients of the parabolic model provide a structured motor-control interpretation. $C_0$ captures a baseline smoothness or motor-cost offset specific to direction and participant. $C_1$ reflects the first-order gain of smoothness with movement duration. $C_2$ captures the curvature of this dependence, whether the smoothness-duration relationship accelerates or decelerates. Critically, $C_2$ was not significantly modified by movement direction or by participant ($p = 0.545$ and $p = 0.266$ respectively). This pattern is compatible with a shared nonlinear temporal structure across directions. In contrast, $C_0$ and, to a lesser extent, $C_1$ differed between directions, indicating a gravity- and anatomy-related shift in the cost of producing smooth movements, not a qualitatively different control regime. The direction-dependent shift in $C_0$ and $C_1$ can be interpreted through the gravitational and anatomical asymmetry of the cervical system. Vertical flexion-extension movements change head orientation relative to gravity, increasing gravitational torque and altering vestibular inflow, whereas horizontal yaw rotations are performed approximately around the gravity axis. The sternocleidomastoid, splenius capitis, and semispinalis capitis contribute differently to rotation, flexion, and extension \citep{Mayoux,Fice18}. Modelling work indicates that cervical extensor moment-generating capacity is generally larger during flexion-extension than during axial rotation \citep{Vasavada,Ackland}, 
and that the head-weight torque increases with head flexion while extensor capacity decreases as posture deviates from neutral \citep{Straker01022009}. Larger torque variations during vertical movements therefore increase jerk-related costs (torque rate scales with jerk as torque scales with acceleration) and produce lower apparent smoothness, consistent with the observed direction effect on $C_0$.

Higher-derivative planning is highly physiologically plausible because, as a consequence of Newton's law, jerk is proportional to the time derivative of force, or yank. Indeed, as $a\sim F$, one must have $j\sim \dot F$. Yank is tightly coupled to neural firing rates and the rapid, synchronized recruitment of motor units during movement initiation \cite{yank0}. Therefore, controlling initial jerk rather than merely initial acceleration maps more directly onto the actual neural commands sent to the muscles, potentially offering a more general and mechanistic framework for motor control.

This interpretation connects to a broader literature on gravity-dependent sensorimotor planning. \cite{PAPAXANTHIS1998103} showed that movement direction relative to gravity modifies the kinematic profiles of vertical arm pointing. \cite{10.1371/journal.pone.0022045} highlighted the temporal asymmetry of upward versus downward arm movements, and \cite{10.7554/eLife.16394} proposed that such direction-dependent kinematics reflect optimal integration of gravitational cues. 
In their review, \cite{doi:10.1152/jn.00381.2019} framed this as a gravitational imprint on sensorimotor 
planning and control, spanning multiple effectors and sensory modalities. The present findings extend 
this logic to the head-neck system, where gravity is not only a mechanical constraint but also modulates 
vestibular and cervical proprioceptive reference frames.


The role of jerk in our framework must not be confounded with the classical minimum-jerk model of \cite{Flash1688}, which explains the smoothness of arm-reaching trajectories by minimizing the time integral of squared jerk. That model is kinematic: It prescribes trajectory shape but does not specify a dynamical mechanism for endpoint realization. Our approach differs in two respects. First, it is formulated as a higher-derivative variational principle (a dynamical principle) rather than a jerk-minimization. Second, jerk enters not only as a smoothness-related descriptor but as an initial dynamical variable that governs endpoint accuracy: In the ideal pointing solution, initial jerk uniquely determines the movement amplitude, while initial acceleration must be close to zero to prevent overshoot or undershoot. In the present model, jerk is not minimized a priori but rather acts as a dynamical planning variable.

Still regarding jerk, we note that the study of \cite{VIKNE2013540} found a relationship between 
the Normalized Jerk Cost (NJC) and average angular speed $\bar v$ during voluntary head movement. 
NJC is another dimensionless index assessing movement smoothness that reads, using our notations, 
NJC$\;=\sqrt{\pi\, I(T)\, \frac{T^5}{x_A^2}}$. As in our model $T=\frac{x_A}{\bar v}$, we find 
NJC$\;=\sqrt{\frac{\omega^5x_A^5}{8\pi}}\, \bar v^{-5/2}$ using the relation $ I =\frac{\cal E}{\omega}$ 
[see (\ref{IT}) below],  (\ref{varEpsilon}) with $a=0$, and (\ref{phidef}). 
The relationship $NJC\sim \bar v^{\alpha}$ has been found by \cite{VIKNE2013540}, with $\alpha\approx 2$ according to a graphical inspection of their results. Our value $-5/2$ is not far from that estimate.

Several limitations should be acknowledged. First, the sample was restricted to young adults aged 18--25, limiting generalization to older populations or patients with neck pain, who may differ in both movement organization and VR tolerance \citep{kim10}. Second, familiarization was conducted in the horizontal condition only, which may have artificially favoured horizontal over vertical performance. Third, the instruction emphasized precision rather than speed; shorter duration therefore cannot be interpreted as better performance in isolation, and this may partly explain why total DidRen time was longer than in prior studies that instructed participants to move as quickly as possible \citep{hage19,DIERICK2024103270}. Fourth, the model assumes a monotonic approach with initial acceleration close to zero, whereas biological movements may include small corrections or submovements. The HD framework should be understood as a principled dynamical approximation rather than a literal trajectory description.

In summary, the present framework aims at providing an alternative to Fitts's law and minimum jerk. 
Fitts's law describes how movement's duration scales with target difficulty. 
While minimum jerk explains smooth trajectories by prescribing a kinematic cost, 
the present HD framework treats head pointing as a smoothness-constrained, higher-derivative control 
problem, in which jerk is a dynamical variable involved in planning and endpoint realization, not merely 
a cost to be reduced. 
The preserved quadratic coefficient $C_2$ across directions points to a shared temporal control law 
for head movements, while direction-dependent $C_0$ and $C_1$ reveal a systematic gravity- and 
anatomy-dependent modulation of motor cost. The recovery of the internal frequency parameter 
$\omega$ from behavioural data alone suggests that this approach may offer a non-invasive 
window into cervical motor planning, with potential applications in the assessment and monitoring of neck 
dysfunction, which should be tested in clinical populations.


\section*{Appendix: Formal developments}\label{formula}

First, we have to compute the integral
\begin{equation}
	I(T)=\frac{1}{2\pi}\int^T \dddot x^2(t)\, dt,
\end{equation}
knowing that $\dddot x(t)=\frac{j}{\cos\phi}\,\cos(\omega\, t+\phi)$ according to (\ref{phidef}) and (\ref{solu1}). One finds
\begin{equation}\label{It}
	I(T)=\frac{j^2}{2\pi\omega\cos^2\!\phi}\left(\frac{\omega\, T}{2}+\frac{1}{2}\cos(\omega\,  T+2\phi)\sin(\omega\, T)\right).
\end{equation}
When $t=T_0=\frac{2\pi}{\omega}$, i.e., when the integral is performed on the whole movement, 
one obtains the relation
\begin{equation}\label{IT}
	I(T_0)=\frac{\cal E}{\omega}\,,
\end{equation}
where $\cal E$ is an invariant of the motion, see \eqref{Invariant} and \eqref{varEpsilon}.

Second, the maximal velocity, $v_M$, is reached at $t^*$ such that 
\begin{equation}
	\ddot x(t^*)=0,\ \dddot x(t^*)<0 \to t^*=-\frac{\phi}{\omega}+\frac{\pi}{\omega}
\end{equation}	
in the case $\phi \geq 0$, leading to 
\begin{equation}\label{vMdef}
	v_M=\dot x(t^*)=\frac{j}{\omega^2}\left(1+\cos\phi\right)=\frac{ x_A}{T_0}\left(1+\cos\phi\right).
\end{equation}
The factor $\frac{ x_A}{T_0}$ is obviously an estimation of the average speed. 

Finally, $LDLJ$, defined by (\ref{LDLJdef}), may be simplified. Indeed, we assume that $\phi$ and $T$ are close to $0$ and $T_0$ for healthy individuals. Therefore we perform a second-order expansion of $LDLJ$ in $T-T_0$ and in $\phi$ using (\ref{It}) and (\ref{vMdef}) to obtain the parabolic expression (\ref{model_eq}), with 
\begin{eqnarray}
	C_0&=&\frac{17}{2}-\ln\left(2\pi^4\right)+4\pi\phi-\frac{15}{2}\phi^2,\\ 
	C_1&=& \frac{\omega}{\pi}\,\left(6+4\pi\phi-5\phi^2\right),\label{C1def}\\
	C_2&=&\frac{\omega^2}{8\pi^2}\,\left(7+8\pi\phi-8\phi^2\right)\label{C2def}\;.
\end{eqnarray}
The coefficients $C_0$, $C_1$ and $C_2$ are strictly positive for small enough values of $\phi$.

\pagebreak
\begin{figure}
	\begin{center}
		\includegraphics[width=0.7\linewidth]{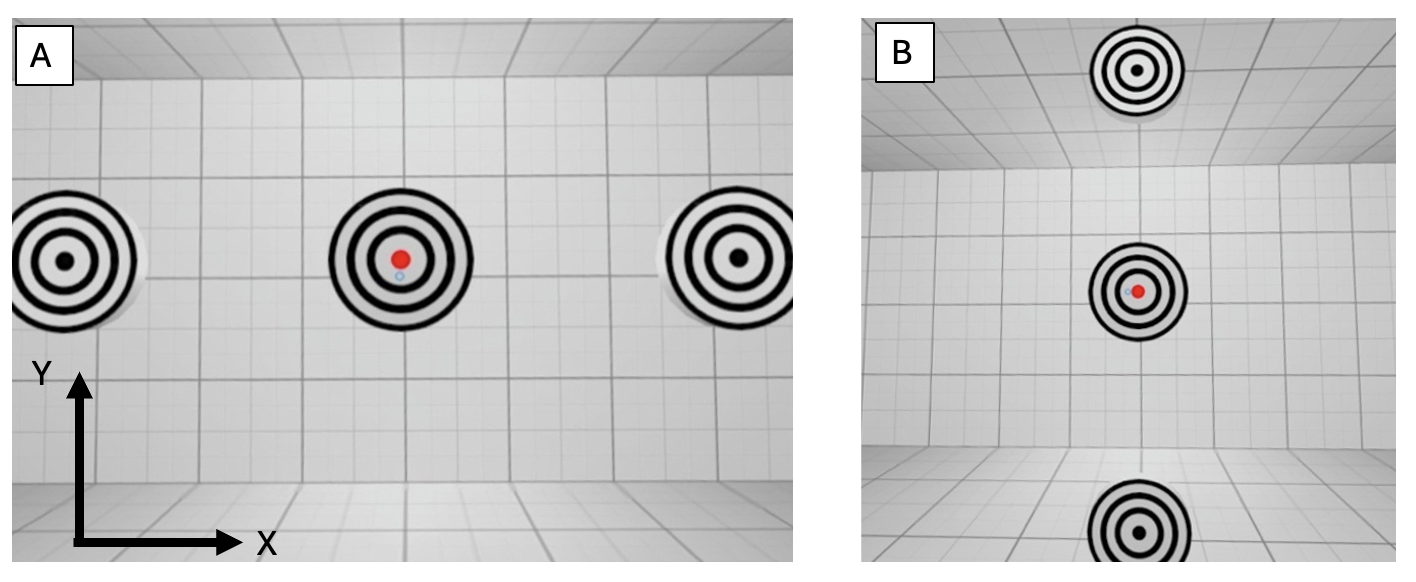}
	\end{center}
	\caption{A: Screenshot of the DidRen VR environment with targets in the horizontal direction. The red zone is the target center used for validation. The laser beam is a blue circle, faintly visible near this zone. B: Same for targets in the vertical direction. A cartesian frame displays the X and Y axis.}
	\label{fig:targets}
\end{figure}


\begin{table}
	\caption{Characteristics of our participants. Data are written under the form mean$\pm$ standard deviation or median [Q1;Q3].}\label{tab:pop}
	\begin{tabular}{c|c}
		\hline
		n & 62 \\
		Sex (M/F) & 38/24 \\
		Age (years)	&21.2$\pm$2.0\\
		NDI (/50) & 3 [1;5] \\
		DidRen time (s) &  44.8$\pm$5.9\\
		\hline
	\end{tabular}\\
	NDI: Neck Disability Index, M: male, F: female.
\end{table}
\begin{table}
	\caption{Comparison of movement duration $T$, accuracy $\delta$, overshoot $\delta_+$, maximal speed $v_M$, and smoothness $LDLJ$, in horizontal (yaw movements) and vertical (pitch movements) directions. Data are written under the form median [Q1;Q3] or mean$\pm$SD, depending on normality. 
	The last two columns show the p-values and paired effect sizes for the Wilcoxon (Cohen's $r$) or paired t-test (Cohen's $d$).}\label{tab:direc}
	\begin{tabular}{c|c|c|c|c}
		\hline
		Parameter & Horizontal &  Vertical & p & $d\, /\, r$ \\
		\hline
		$T$ (s)&  1.364 [1.253; 1.542] & 1.261 [1.223; 1.525] & 0.850 & -0.025 \\
		$\delta$ ($^{\rm o}$) &  1.135 [1.054; 1.281] & 1.221 [1.113; 1.362] & 0.147 & -0.184  \\
		$\delta_+$ ($^{\rm o}$)& 0.531 [0.175; 0.730] & 0.358 [0.175; 0.730] & 0.001 & 0.413 \\
		$v_M$ ($^{\rm o}/$s) &  72.3 [58.1; 88.4] & 65.9 [54.5; 82.1] & 0.025 & 0.284 \\
		$LDLJ$ &  -9.286$\pm$0.633 & -9.935$\pm$0.669 & $<0.001$ & 0.996 \\
		\hline
	\end{tabular}
	
\end{table}

\end{document}